\documentclass[fleqn]{elsart}
\usepackage{amsmath}

\def\b{\begin{equation}}
\def\e{\end{equation}}

\def\ba{\begin{eqnarray}}
\def\ea{\end{eqnarray}}

\begin{document}

\begin{frontmatter}

\title{A note on the  Painlev\'e analysis of a $\mathbf (2+1)$ dimensional
Camassa-Holm equation}

\author{P. R. Gordoa\thanksref{label1}},
\author{A. Pickering\thanksref{label1}}, and
\author{M. Senthilvelan\thanksref{label2}\corauthref{cor1}}
\ead{senthilvelan@cnld.bdu.ac.in}
\corauth[cor1]{Corresponding Author}
\address[label1]{Area de Matem\'atica Aplicada, ESCET, Universidad Rey 
Juan Carlos, C/~Tulip\'an s/n, 28933 M\'ostoles, 
Madrid, Spain}
\address[label2]{Centre for Nonlinear Dynamics, Department of Physics,
Bharathidasan University, Tiruchirappalli - 620 024, India}


\date{}

\begin{abstract}

We investigate the Painlev\'{e} analysis for a $(2+1)$ dimensional 
Camassa-Holm equation. Our results show that it admits only 
weak Painlev\'{e} expansions. This then confirms the limitations of 
the Painlev\'e test as a test for complete integrability when applied
to non-semilinear partial differential equations.

\end{abstract}


%
%



\end{frontmatter}


\section{Introduction}

Recently there has been a great deal of interest in the study of 
the integrability 
properties of the Camassa-Holm (CH) equation,
\begin{eqnarray} 
u_t+2\kappa u_x-u_{xxt}+3uu_x = 2u_xu_{xx}+uu_{xxx},
\label{11ch}
\end{eqnarray} 
where subscripts denote partial derivatives   and $\kappa$ is a constant
\cite{CamHol:1}.  Equation (\ref{11ch})  was first derived by Fokas and 
Fuchssteiner using the method of recursion operators \cite{Fok}.  However, 
it became more widely known when it was derived from physical considerations 
as a water wave equation by Camassa  and Holm, by using asymptotic expansions 
directly in the Hamiltonian for  Euler's equations governing inviscid 
incompressible flow in the shallow  water regime \cite{CamHol:1}.  
They showed that equation (\ref{11ch}) is completely integrable  for 
arbitrary values of $\kappa$, and that for $\kappa=0$ it admits a special 
kind of travelling wave solution, the peakon, of the form $ce^{-|x-ct|}$. 
Further, they  showed that it possesses a  Lax pair and bi-Hamiltonian 
structure.  Subsequently, several works have  been devoted to studying 
the underlying mathematical and physical properties  of equation (\ref{11ch})
\cite{Coop,GilPic,Fuch,Senthil,Sch,Fish,Cons,Liu,Pick:1,Camassa1,Dullin}.

Interestingly, several higher dimensional generalizations of the CH equation 
have also been proposed.  For example, Kraenkel and Zenchuk
constructed the following $(2+1)$ dimensional 
integrable generalization of the CH equation \cite{kraenkel:1},
\ba
& &  m_t-um_x-2mu_x  =   0,\nonumber\\
& & mu_y+u_x-u_{xx}-2p_x  =  0,\nonumber\\
& &  m^2p_y+mp_x+mp_{xx}-pm_x-m_xp_x  =  0,
\label{21ch}
\ea
where $m$, $u$ and $p$ are functions of $(x,y,t)$. In \cite{kraenkel:1},
equation (\ref{21ch}) was derived from a system of $(2+1)$ 
dimensional shallow water equations by using a multiscale decomposition, 
and was also shown to have a Lax pair. The spectral problem for this equation  
was solved by Zenchuk \cite{Zen1}. Kraenkel et al.\ \cite{kraenkel:2}
studied the invariance properties of (\ref{21ch}) through Lie symmetry 
analysis and explored its similarity reductions, giving Lax pairs for all 
the resulting $(1+1)$ dimensional partial differential equations (PDEs) 
and thus establishing their integrability.  
Recently, Johnson has derived another $(2+1)$ dimensional CH equation, 
which has a structure similar to that of the Kadomtsev-Petviashvili equation, 
again within the context of equations governing water waves \cite{Johnson:1}.
The integrability of this equation is discussed in \cite{GPS2004}.

Even though many interesting mathematical properties have been studied 
for the $(2+1)$ dimensional CH equation (\ref{21ch}), an important question 
remains unanswered, that is, whether or not it passes the Painlev\'{e} (P-) 
test \cite{WTC1,WTC2}. Indeed, certain difficulties arise when applying 
the P-test to equation (\ref{21ch}).
These difficulties are similar to those that arise for the class of equations
studied in \cite{GilPic}, and which also arise for the special case of the CH
equation itself. However, in \cite{GilPic} it was shown how these difficulties
can be overcome by including an extra lower order term in the Painlev\'e
expansion; in particular, it was shown that the CH equation admits only weak
Painlev\'e expansions. We find that a similar modification of the leading 
order analysis allows us to apply the P-test to equation (\ref{21ch}). We 
find that this equation admits only weak Painlev\'e expansions.

\section{Painlev\'{e} analysis}

In this section we apply the P-test  \cite{WTC1,WTC2} to the $(2+1)$ 
dimensional CH equation (\ref{21ch}).    
The P-test consists essentially of three steps: 
(i) determination of leading-order behaviours, (ii) identifying the
resonances, and (iii) checking the corresponding compatibility conditions
(in order to avoid logarithmic branching).

\subsection{(i) Leading order analysis}

As a first step we consider seeking local Laurent expansions about
a noncharacteristic movable singular manifold $\varphi=0$. We find
that the leading order exponent for $u$ cannot be negative, and so
assume, following \cite{GilPic}, that the leading orders of the 
solutions of equation (\ref{21ch}) are
\b
u  \sim c \psi_t+u_0\phi^\alpha,\qquad
m  \sim m_0\phi^\beta,\qquad
p  \sim p_0\phi^\gamma,
\label{lead1}
\e
where $c$ is a constant to be determined. We note that
here we are using Kruskal's ansatz \cite{Krus}, that is, we are taking 
$\phi = x+\psi(y,t)$ and all coefficients in our Laurent expansions 
(for example, $u_0$, $m_0$ and $p_0$ above) to be functions of $(y,t)$ only.

Balancing the most dominant terms, we find that $c=1$, and are led to the 
following two cases: 

\bigskip
\noindent {\underline{Case 1}} \quad  $\alpha=\frac{1}{2}$, $\beta=-1$, 
$\gamma=-\frac{1}{2}$ with 
$m_0 = \frac{1}{2\psi_y}$, $p_0=-\frac{u_0}{2}$, $u_0$ arbitrary.
\medskip

\noindent 
{\underline{Case 2}} \quad $\alpha=\frac{1}{2}$, $\beta=-1$, 
$\gamma=\frac{1}{2}$ 
with $m_0 = -\frac{1}{2\psi_y},$ $u_0$ and $p_0$ arbitrary.

\subsection{(ii) Resonances}
The next step in the P-test is to find the resonances, that is,  
the powers at which arbitrary functions appear in the series.  
As we have encountered two branches in the leading order analysis, 
we present the calculations for each case separately.

\bigskip
\noindent
{\underline{Case 1}}
\medskip

Substituting the expressions 
\ba
& & u  = \psi_t+u_0\phi^{\frac{1}{2}}+u_r\phi^{r+\frac{1}{2}},\nonumber\\
& & m  = m_0\phi^{-1}+m_r\phi^{r-1},\nonumber\\
& & p  = p_0\phi^{-\frac{1}{2}}+p_r\phi^{r-\frac{1}{2}},
\label{reson1}
\ea
into the dominant terms tells us that the recursion relation for the
coefficients of our Laurent expansions is of the form
\b
\left(\begin{array}{ccc}
\frac{r}{\psi_y} & ru_0 &  0 \\
-\frac{1}{2}(2r+1)(r-1) & \frac{1}{2}u_0\psi_y &  1-2r \\
 0 & -\frac{1}{8}(2r-1)u_0  &\frac{r(2r-1)}{4\psi_y}
\end{array}\right)
\left(\begin{array}{c}
u_r\\
m_r\\
p_r
\end{array}\right)
 = 
\left(\begin{array}{c}
f_r\\g_r\\h_r
\end{array}\right),
\label{reson2}
\e
where as usual $f_r$, $g_r$ and $h_r$ are functions of $\psi$, previous
coefficients, and derivatives thereof. We thus obtain the resonances as
$r = -1,0,\frac{1}{2},\frac{1}{2},1$. The resonance $r=-1$ is associated 
with the arbitrariness of the function $\psi$; $r=0$ corresponds to the 
function $u_0$ being arbitrary. The double resonance $r=\frac{1}{2}$ 
indicates that we must modify our Laurent series in order to include 
half-integer powers of $\phi$, thus obtaining a Puiseux series, and that 
two of the coefficients
$(u_{\frac{1}{2}},m_{\frac{1}{2}},p_{\frac{1}{2}})$ therein will be arbitrary.
Finally the resonance $r=1$ indicates that 
one of the coefficients $(u_1,m_1,p_1)$ will be arbitrary.

\bigskip
\noindent
{\underline{Case 2}} 
\medskip

In a similar way, substituting the expressions 
\ba
& & u  = \psi_t+u_0\phi^{\frac{1}{2}}+u_r\phi^{r+\frac{1}{2}},\nonumber\\
& & m  = m_0\phi^{-1}+m_r\phi^{r-1},\nonumber\\
& & p  = p_0\phi^{\frac{1}{2}}+p_r\phi^{r+\frac{1}{2}},
\ea
into the dominant terms of (\ref{21ch}) tells us that the form of the 
recursion relation is
\b
\left(\begin{array}{ccc}
-\frac{r}{\psi_y} & ru_0 &  0 \\
-r(r+\frac{1}{2}) & \frac{1}{2}u_0\psi_y &  0 \\
0 & -\frac{p_0}{2}(r+\frac{1}{2}) &  -\frac{r}{2\psi_y}(r+\frac{1}{2}) 
\end{array}\right)
\left(\begin{array}{c}
u_r \\
m_r \\
p_r
\end{array}\right)
 = 
\left(\begin{array}{c}
\tilde f_r\\ \tilde g_r\\ \tilde h_r
\end{array}\right),
\e
where the functions $\tilde f_r$, $\tilde g_r$ and $\tilde h_r$ are 
functions of $\psi$, previous coefficients, and derivatives thereof.
We thus obtain the resonances $r = -1,-\frac{1}{2},0,0,\frac{1}{2}$.
The double resonance $r=0$ corresponds 
to the arbitrariness of the functions $u_0$ and $p_0$. The  resonance
$r=\frac{1}{2}$ indicates once again that we need to modify our Laurent 
series in order to include half-integer powers of $\phi$, and that one of 
the coefficients $(u_{\frac{1}{2}},m_{\frac{1}{2}},p_{\frac{1}{2}})$ in
the resulting series will be arbitrary.
We might expect that the resonance $r=-\frac{1}{2}$ means that we need
to use the perturbative Painlev\'e test. However, we will see that this
resonance, on this occasion, can be accommodated much more easily.

We note that, since we have non-integer resonances (in fact, in each of the 
above cases), equation (\ref{21ch}) does not pass the Painlev\'e test. It
remains to check whether or not it admits weak Painlev\'e --- or Puiseux ---
expansions (see \cite{WTC2}, also \cite{GilPic,AP}).

\subsection{(iii) Checking compatibility conditions}

We now check the compatibility conditions, using the following expansions.

\bigskip
\noindent
{\underline{Case 1}}
\medskip

We substitute
\b
u = \psi_t+\phi^{\frac{1}{2}}\sum_{j=0}^{\infty}u_{\frac{j}{2}}
\phi^{\frac{j}{2}},\;\;\;
m = \phi^{-1}\sum_{j=0}^{\infty}m_{\frac{j}{2}}
\phi^{\frac{j}{2}},\;\;\;
p = \phi^{-\frac{1}{2}}\sum_{j=0}^{\infty}p_{\frac{j}{2}}
\phi^{\frac{j}{2}},
\label{res1}
\e
with $\phi = x+\psi(y,t)$ and coefficients $u_i=u_i(y,t)$, $m_j=m_j(y,t)$ 
and $p_k=p_k(y,t)$, and with
$m_0=\frac{1}{2\psi_y}$, $p_0=-\frac{u_0}{2}$ and $u_0$ arbitrary,
into equation (\ref{21ch}).  

At the point of determining the coefficients 
$(u_{\frac{1}{2}},m_{\frac{1}{2}},p_{\frac{1}{2}})$, we find that
$p_{\frac{1}{2}}$ is arbitrary, and that one of $u_{\frac{1}{2}}$
and $m_{\frac{1}{2}}$ is also arbitrary, with both compatibility 
conditions satisfied. We choose to leave $m_{\frac{1}{2}}$ arbitrary,
solving for $u_{\frac{1}{2}}$ as
\b
u_{\frac{1}{2}} = -\frac{\psi_{yt}}{\psi_y} - u_0m_{\frac{1}{2}}\psi_y.
\label{res3}
\e
At the point of determining the coefficients 
$(u_1,m_1,p_1)$, we find that any one of these three can be left 
arbitrary with satisfied compatibility condition. Thus for example
we may choose $m_1$ to be arbitrary, and solve for $u_1$ and $p_1$.
All subsequent coefficients are then determined in terms of the
five arbitrary functions $(\psi,u_0,m_{\frac{1}{2}},p_{\frac{1}{2}},m_1)$.  
Since all compatibility conditions are satisfied, we see that in this
case, equation (\ref{21ch}) admits a weak Painlev\'e expansion. 

\bigskip
\noindent
{\underline{Case 2}}
\medskip

In this case, instead of the series that would usually be chosen to
check compatibility conditions, we substitute the series
\b
u = \psi_t+\phi^{\frac{1}{2}}\sum_{j=0}^{\infty}u_{\frac{j}{2}}
\phi^{\frac{j}{2}},\;\;\;
m = \phi^{-1}\sum_{j=0}^{\infty}m_{\frac{j}{2}}
\phi^{\frac{j}{2}},\;\;\;
p = p_{-\frac{1}{2}}+\phi^{\frac{1}{2}}\sum_{j=0}^{\infty}p_{\frac{j}{2}}
\phi^{\frac{j}{2}},
\label{res6}
\e
with $m_0=-\frac{1}{2\psi_y}$, and $u_0$ and $p_0$ arbitrary, into 
equation (\ref{21ch}).
The difference with the usual series that would be taken lies in the inclusion
of the extra term $p_{-\frac{1}{2}}$. It is the inclusion of this extra term
that allows us to accommodate the resonance $r=-\frac{1}{2}$, since we find
that $p_{-\frac{1}{2}}$ is left arbitrary.

At the point of determining the coefficients 
$(u_{\frac{1}{2}},m_{\frac{1}{2}},p_{\frac{1}{2}})$, we find that
any one of these three can be left arbitrary with satisfied compatibility 
condition. Thus for example we may choose $m_{\frac{1}{2}}$ to be arbitrary, 
and solve for $u_{\frac{1}{2}}$ and $p_{\frac{1}{2}}$.
All subsequent coefficients are then determined in terms of the
five arbitrary functions $(\psi,p_{-\frac{1}{2}},u_0,p_0,m_{\frac{1}{2}})$.  
Since all compatibility conditions are satisfied, we see that also in this
case, equation (\ref{21ch}) admits a weak Painlev\'e expansion. 
\section{Conclusions}
In this letter we have carried out a detailed investigation of the 
Painlev\'{e} analysis of the completely integrable $(2+1)$ dimensional 
generalization of the CH equation, equation (\ref{21ch}).  We have 
used a modified version of the leading order analysis, as adopted in 
\cite{GilPic}. In this way, we have shown that the equation under study 
admits only weak Painlev\'e expansions. This example therefore confirms 
the limitation of the Painlev\'{e} test as a test for complete integrability
when applied to non-semilinear PDEs.   
\section*{Acknowledgements}
 
The work of MS forms part of a Department of Science and Technology,
Government of India sponsored research project.
The work of PRG and AP is supported in part by the DGESYC
under contract BFM2002-02609, and that of AP by the Junta de Castilla y
Le\'on under contract SA011/04. PRG currently holds a Ram\'on y Cajal research
fellowship awarded by the Ministry of Science and Technology of Spain,
which support is gratefully acknowledged.  This work was carried out
during the visit of MS to the University of Salamanca, funded under
contract BFM2002-02609, in June 2003.


\end{document}